\documentclass[pra,a4paper,tightenlines,notitlepage,final,twocolumn,showpacs,preprint,footinbib,10pt,bibnotes]{revtex4}
\usepackage{amssymb}

\usepackage{amsmath}

\input{tcilatex}

\begin{document}

\title{Strategies for Estimating Quantum Lossy Channels}
\author{Xian-Ting Liang\thanks{%
E-mail address: xtliang@ustc.edu}}
\affiliation{Department of Physics, Ningbo University, Ningbo, Zhejiang 315211, China}

\begin{abstract}
Due to the anisotropy of quantum lossy channels one must choose optimal
bases of input states for best estimating them. In this paper, we obtain
that the equal probability Schr\"{o}dinger cat states are optimal for
estimating a single lossy channel and they are also the optimal bases of
input states for estimating composite lossy channels. On the other hand, by
using the symmetric logarithmic derivative (SLD) Fisher information of
output states exported from the lossy channels we obtain that if we take the
equal probability Schr\"{o}dinger cat states as the bases of input states
the maximally entangled inputs are not optimal, however if the bases of the
input states are not the equal probability Schr\"{o}dinger cat states the
maximally entangled input states may be optimal for the estimating composite
lossy channel.
\end{abstract}

\maketitle

\section{Introduction}

A quantum noisy channel can be expressed with a trace-preserving completely
positive map: $\varepsilon :\rho \rightarrow \rho ^{\prime }\left( \vec{%
\varsigma}\right) .$ Here, $\rho $ and $\rho ^{\prime }\left( \vec{\varsigma}%
\right) $ are density matrixes in the Hilbert space $\mathcal{H}$, and $\vec{%
\varsigma}=\left( \varsigma _{1},\varsigma _{2},...\varsigma _{n}\right) \in
\Gamma $ are parameters characterizing the channel. So a single parameter
quantum noisy channel can be expressed as%
\begin{equation}
\rho ^{\prime }\left( \varsigma \right) =\varepsilon \left( \rho \right) .
\label{e1}
\end{equation}%
Due to the complete positivity of the map, it can be expanded to composite
quantum systems in $\mathcal{H\otimes H}$. The composite channels in the
expanded systems have two forms, one is%
\begin{equation}
\varrho ^{\prime }\left( \varsigma \right) =\varepsilon \otimes I\left(
\varrho \right) ,  \label{e2}
\end{equation}%
called \emph{mixed} noisy channel; the other is%
\begin{equation}
\varrho ^{\prime \prime }\left( \varsigma \right) =\varepsilon \otimes
\varepsilon \left( \varrho \right) ,  \label{e3}
\end{equation}%
called \emph{double} noisy channel. Here, $\varrho ,\varrho ^{\prime
},\varrho ^{\prime \prime }\ $are density matrixes in $\mathcal{H\otimes H}$%
, and $I,$ $\varepsilon $ are the single quantum identity and noisy channel.

For a known quantum noisy channel at lest two subjects are interested. One
is to determine its information capacities, the best probability to
understand the input state under assumption that the action of the quantum
channel is known. Although the capacities of quantum noisy channels have not
been solved thoroughly much effort has been put into this topic and many
results are obtained \cite{capacities}. Another topic, the estimation of
quantum noisy channel has also been attracted much attention in last years %
\cite{estimation02} \cite{estimation03} \cite{Fujiwara042304} because it is
also important in quantum information theory. The estimation of quantum
noisy channel is to identify a quantum noisy channel as the type of the
channel is known but its quality is unknown. The quality of the channel can
be characterized with some parameters. Thus estimating some channel is equal
to estimating its certain parameters, which may be appealed to the quantum
estimation theory \cite{estimation01}.

Quantum estimation theory is one about seeking the best strategy for
estimating one or more parameters of a density operator of a quantum
mechanical system. About how to estimate a quantum noisy channel we refer
the readers to the Refs. \cite{estimation01} and recent \cite{estimation02}.
In this paper we restrict our attention in two aspects of the estimation of
a quantum noisy channel, \emph{lossy channel }(which will be described in
section II). Because of the anisotropy of the quantum lossy channels,
different input states must have different effects for estimating the
quantum noisy channel. So at first, we will discuss what \emph{coherent
states} are optimal for estimating the single quantum lossy channel and what
bases of the input states are optimal for estimating the composite lossy
channels? Because entanglement has been taken a kind of resource for
processing quantum information, secondly, we will discuss: can the
estimation of the composite channels $\varepsilon \otimes I\left( \varrho
\right) $ and $\varepsilon \otimes \varepsilon \left( \varrho \right) $ be
improved by using entangled input states? This paper is constructed as
follows. In section II we shall set up a model of quantum lossy channel and
explain why we choose coherent states to estimate these channels. In section
III we shall seek for the optimal input coherent states or optimal bases of
input states for estimating the single and the composite lossy channels. In
section IV we shall calculate the symmetric logarithmic derivative (SLD)
Fisher information of output states exported from the channels $\varepsilon $%
, $\varepsilon \otimes I$ and $\varepsilon \otimes \varepsilon $ and answer
whether the entangled input states improve the estimation. A brief
conclusion will close this paper in last section.

\section{Lossy channel and Schr\"{o}dinger cat state}

We set the lossy channel to be estimated is described by the following
physical model. A quantum system, such as photons in state $\rho $ is in a
vacuum environment, the evolution of the state is a completely positive map: 
$\rho ^{\prime }=\varepsilon \left( \rho \right) $. In this model, making
use of the language of master equation we can obtain that the interaction of
the in question system with its environment makes the system evolving
according to 
\begin{eqnarray}
\frac{\partial \rho }{\partial \tau } &=&\hat{J}\rho +\hat{L}\rho ;\hat{J}%
\rho =\eta \sum_{i}a_{i}\rho a_{i}^{\dagger }  \notag \\
\hat{L}\rho  &=&-\sum_{i}\frac{\eta }{2}\left( a_{i}^{\dagger }a\rho +\rho
a_{i}^{\dagger }a\right) ,  \label{e4}
\end{eqnarray}%
where $\eta $ is the energy decay rate. The formal solution of Eq.(\ref{e4})
may be written as%
\begin{equation}
\rho \left( \tau \right) =\exp \left( \left( \hat{J}+\hat{L}\right) \tau
\right) \rho \left( 0\right) ,  \label{e5}
\end{equation}%
which leads to the solution for the initial single-mode $\left| \alpha
\right\rangle \left\langle \beta \right| $ 
\begin{equation}
\exp \left[ \left( \hat{J}+\hat{L}\right) \tau \right] \left| \alpha
\right\rangle \left\langle \beta \right| =\left\langle \beta \right| \left.
\alpha \right\rangle ^{t^{2}}\left| \alpha t\right\rangle \left\langle \beta
t\right| ,  \label{e6}
\end{equation}%
where $t=e^{-\frac{1}{2}\eta \tau }.$ Estimating this channel is equal to
estimating the parameter $\eta .$

As known, in order to estimate the channel, one, for example Alice must
prepare many identical initial states, input states $\rho $ and another one,
for example Bob must measure the output samples exported from this channel.
For enhancing the detection efficiency we use coherent states to be the
input states \cite{Jeongetal052308}. Because we do not know what coherent
state is the optimal input state for the estimation in advance, we generally
set this state be the superposition of coherent states $\left| \alpha
\right\rangle $ and $\left| -\alpha \right\rangle $, namely, a
Schr\"{o}dinger cat state%
\begin{equation}
\left| \varphi \right\rangle =\mathcal{A}\left| \alpha \right\rangle +%
\mathcal{B}\left| -\alpha \right\rangle .  \label{e7}
\end{equation}%
This state is considered one of realizable mesoscopic quantum systems \cite%
{YurkeetalPRL1986}. Zheng \cite{Zheng} has shown the method for preparing
this state and the measurement scheme of this state has been given in \cite%
{Jeong01}. Set $\left| \alpha \right| \gg 1$ (in fact only if $\left| \alpha
\right| \geq 3$)$,$ then $\left\langle \alpha \right| \left. -\alpha
\right\rangle \simeq 0.$ Thus, $\left| \alpha \right\rangle $ and $\left|
-\alpha \right\rangle $ can be taken into a pair orthogonal bases. Setting $%
\mathcal{A}\approx \sin \theta ,$ $\mathcal{B}\approx \cos \theta $ we have%
\begin{equation}
\left| \varphi \right\rangle =\sin \theta \left| \alpha \right\rangle +\cos
\theta \left| -\alpha \right\rangle .  \label{e8}
\end{equation}%
In the time-varying bases $\left[ \left| \alpha _{1}t\right\rangle ,\left|
\alpha _{2}t\right\rangle \right] ^{T},$ the state $\rho =\left| \varphi
\right\rangle \left\langle \varphi \right| $ passing through the lossy
channel becomes 
\begin{eqnarray}
\rho ^{\prime } &=&\varepsilon \left( \rho \right) \approx \left( 
\begin{array}{cc}
\sin ^{2}\theta  & \sin \theta \cos \theta e^{-2\left| \alpha \right|
^{2}\eta \tau } \\ 
\sin \theta \cos \theta e^{-2\left| \alpha \right| ^{2}\eta \tau } & \cos
^{2}\theta 
\end{array}%
\right)   \notag \\
&=&\frac{1}{2}I+\chi \sin \theta \cos \theta \sigma _{x}+\left( \sin
^{2}\theta -\frac{1}{2}\right) \sigma _{z},  \label{e9}
\end{eqnarray}%
where $\chi =e^{-2\left| \alpha \right| ^{2}\eta \tau },$ $\sigma _{i}$ $%
\left( i=x,\text{ }y,\text{ }z\right) $ are Pauli matrixes and $\left\langle
\alpha \right| \left. -\alpha t\right\rangle \rightarrow 0$ is set (namely,
we set the time of the system evolution is very short, $\tau \rightarrow 0,$ 
$t\rightarrow 1$). We call the map of Eq.(\ref{e9}) the single lossy channel.

\section{Optimal input states and optimal bases of input states for
estimating the lossy channels}

In this section we answer the first question put forward in section II,
namely, what (coherent) input states are optimal for estimating the single
lossy channel and what bases of the (coherent) input states are optimal for
estimating the composite lossy channels? In the following we will firstly
investigate the single lossy channel then the mixed lossy channel $%
\varepsilon \otimes I$ and the double lossy channel $\varepsilon \otimes
\varepsilon $.

In the quantum estimation theory, in order to estimate a channel, at first,
one must introduce a cost function. In general, delta function is chosen for
this aim, namely,%
\begin{equation}
C\left( \hat{\zeta},\zeta \right) =-\prod_{i=1}^{m}\delta \left( \hat{\zeta}%
_{i}-\zeta _{i}\right) ,  \label{e10}
\end{equation}%
where $\hat{\zeta}_{i}$ called estimators which are always a function of
observing data and it describes the strategy for calculating the estimates; $%
\zeta _{i}$ are parameters to be estimated. It is given that the optimal
estimation is to seek the POVM generators $d\Pi _{m}\left( \zeta \right) $
for which%
\begin{equation}
\left\{ 
\begin{array}{c}
\left( \Upsilon -W_{m}\left( \zeta \right) \right) d\Pi _{m}\left( \zeta
\right) =0, \\ 
\Upsilon -W_{m}\geqslant 0.%
\end{array}%
\right.   \label{e11}
\end{equation}%
Here, 
\begin{equation}
W_{m}\left( \zeta \right) =Z\left( \zeta \right) \rho _{m}\left( \zeta
\right) ,  \label{e12}
\end{equation}%
where $Z\left( \zeta \right) $ is the prior probability density function
(PDF), and 
\begin{equation}
\Upsilon =\tsum_{m}\Upsilon _{m}=\tsum_{m}\int_{\Theta }W_{m}\left( \zeta
\right) d\Pi _{m}\left( \zeta \right) ,\zeta \in \Gamma .  \label{e13}
\end{equation}%
In our problem, we shall find out a optimal input state, namely a optimal
angle $\theta $ in Eq.(\ref{e9}), where $0<\chi \leq 1$ is supposed. In the
following subsection A, we shall investigate what state is the optimal input
state for estimating the single lossy channel. In subsection B, we shall
investigate that when we use two-mode entangled state estimating the
composite lossy channels (include mixed lossy channel and double lossy
channel), if we measure the output state separately, what bases of the input
states are optimal. In subsection C, we shall look for the optimal bases of
input states for estimating composite channels as we measure the output
states jointly.

\subsection{Single lossy channel}

The schematic diagram for estimating the single lossy channel can be
expressed as Fig.1 (above part).\FRAME{ftbphFU}{229.875pt}{110.3125pt}{0pt}{%
\Qcb{Schematic diagrams for estimating the single lossy channel (above) and
composite lossy channel through separately measuring the out states (blow)}}{%
}{Figure}{\special{language "Scientific Word";type
"GRAPHIC";maintain-aspect-ratio TRUE;display "USEDEF";valid_file "T";width
229.875pt;height 110.3125pt;depth 0pt;original-width
226.625pt;original-height 107.625pt;cropleft "0";croptop "1";cropright
"1";cropbottom "0";tempfilename 'HOWX6H05.wmf';tempfile-properties "XPR";}}%
Because there is no prior knowledge about the angle $\theta $, we assign to
it a uniform prior probability density function, namely%
\begin{equation}
Z\left( \chi \right) =\frac{1}{2\pi }.  \label{e14}
\end{equation}%
The POVM generator is 
\begin{equation}
d\Pi \left( \chi \right) =\frac{1}{2\pi }\left( I+\chi \sigma _{x}\sin
2\theta -\sigma _{z}\cos 2\theta \right) d\theta ,  \label{e15}
\end{equation}%
and 
\begin{equation}
W\left( \chi \right) =\frac{1}{4\pi }\left( I+\chi \sigma _{x}\sin 2\theta
-\sigma _{z}\cos 2\theta \right) .  \label{e16}
\end{equation}%
So we have%
\begin{equation}
\Upsilon -W\left( \chi \right) =\frac{1}{8\pi }\left[ \left( 1+\chi
^{2}\right) I-2\chi \sigma _{x}\sin 2\theta +2\sigma _{z}\cos 2\theta \right]
.  \label{e17}
\end{equation}%
The eigenvalues of $\Upsilon -W\left( \chi \right) $ are 
\begin{equation}
\lambda =1+\chi ^{2}\pm 2\sqrt{\cos ^{2}2\theta +\chi ^{2}\sin ^{2}2\theta }.
\label{e18}
\end{equation}%
Because $0<\chi \leq 1,$ if and only if $\theta =\pm \pi /4,$ $\pm 3\pi /4$
we have%
\begin{equation}
\left\{ 
\begin{array}{c}
\left[ \Upsilon -W\left( \chi \right) \right] d\Pi \left( \chi \right) =0,
\\ 
\Upsilon -W\left( \chi \right) \geqslant 0.%
\end{array}%
\right.   \label{e19}
\end{equation}%
It shows that the equal probability Schr\"{o}dinger cat states $\left|
\varphi \right\rangle =\left( \left| \alpha \right\rangle \pm \left| -\alpha
\right\rangle \right) /\sqrt{2}$ are optimum for estimating the lossy
channel $\varepsilon $.

\subsection{Composite lossy channels (separated measurements)}

In the following, we shall find out the optimal bases of the input states
for estimating the mixed lossy channel as we separately measure the out
states exported from composite lossy channels. The schematic diagram is
shown in Fig.1 (below part). If the estimated channel is a mixed channel we
can use possibly entangled states as their inputs. Set the input state be $%
\varrho =\left| \Phi \right\rangle \left\langle \Phi \right| $, where $%
\left| \Phi \right\rangle \in \mathcal{H}_{1}\otimes \mathcal{H}_{2}$. By
the Schmidt decomposition, the vector $\left| \Phi \right\rangle $ is
represented as%
\begin{equation}
\left| \Phi \right\rangle =\sqrt{\gamma }\left| \varphi _{1}\right\rangle
\left| \varphi _{2}\right\rangle +\sqrt{1-\gamma }\left| \psi
_{1}\right\rangle \left| \psi _{2}\right\rangle ,  \label{e20}
\end{equation}%
where $\gamma $ is a real number between $0$ and $1$, and $\left\{ \varphi
_{1},\psi _{1}\right\} $ and $\left\{ \varphi _{2},\psi _{2}\right\} $ are
orthonormal bases of $\mathcal{H}_{1}=C^{2}$ and $\mathcal{H}_{2}=C^{2}.$ We
generally set they are 
\begin{eqnarray}
\left| \varphi _{i}\right\rangle  &=&\sin \theta \left| \alpha \right\rangle
+\cos \theta \left| -\alpha \right\rangle ,  \notag \\
\left| \psi _{i}\right\rangle  &=&\cos \theta \left| \alpha \right\rangle
-\sin \theta \left| -\alpha \right\rangle .  \label{e21}
\end{eqnarray}%
From Eq.(\ref{e20}) we have%
\begin{eqnarray}
\varrho  &=&\left| \Phi \right\rangle \left\langle \Phi \right| =\gamma
\left| \varphi _{1}\right\rangle \left\langle \varphi _{1}\right| \otimes
\left| \varphi _{2}\right\rangle \left\langle \varphi _{2}\right|   \notag \\
&&+\sqrt{\gamma \left( 1-\gamma \right) }\left| \varphi _{1}\right\rangle
\left\langle \psi _{1}\right| \otimes \left| \varphi _{2}\right\rangle
\left\langle \psi _{2}\right|   \notag \\
&&+\sqrt{\gamma \left( 1-\gamma \right) }\left| \psi _{1}\right\rangle
\left\langle \varphi _{1}\right| \otimes \left| \psi _{2}\right\rangle
\left\langle \varphi _{2}\right|   \notag \\
&&+\left( 1-\gamma \right) \left| \psi _{1}\right\rangle \left\langle \psi
_{1}\right| \otimes \left| \psi _{2}\right\rangle \left\langle \psi
_{2}\right| .  \label{e22}
\end{eqnarray}%
As the state $\varrho $ pass through the mixed channel it becomes 
\begin{eqnarray}
\varrho ^{\prime } &=&\varepsilon \otimes I\left( \varrho \right)   \notag \\
&=&\gamma \varepsilon \left( \left| \varphi _{1}\right\rangle \left\langle
\varphi _{1}\right| \right) \otimes I\left( \left| \varphi _{2}\right\rangle
\left\langle \varphi _{2}\right| \right)   \notag \\
&&+\sqrt{\gamma \left( 1-\gamma \right) }\varepsilon \left( \left| \varphi
_{1}\right\rangle \left\langle \psi _{1}\right| \right) \otimes I\left(
\left| \varphi _{2}\right\rangle \left\langle \psi _{2}\right| \right)  
\notag \\
&&+\sqrt{\gamma \left( 1-\gamma \right) }\varepsilon \left( \left| \psi
_{1}\right\rangle \left\langle \varphi _{1}\right| \right) \otimes I\left(
\left| \psi _{2}\right\rangle \left\langle \varphi _{2}\right| \right)  
\notag \\
&&+\left( 1-\gamma \right) \varepsilon \left( \left| \psi _{1}\right\rangle
\left\langle \psi _{1}\right| \right) \otimes I\left( \left| \psi
_{2}\right\rangle \left\langle \psi _{2}\right| \right) .  \label{e23}
\end{eqnarray}%
If Bob samples the data by separate measuring the out states from channel $%
\varepsilon _{1}$ OR channel $\varepsilon _{2}$, say channel $\varepsilon
_{1}$, means%
\begin{eqnarray}
\varrho _{1}^{\prime } &=&tr_{2}\left( \varrho ^{\prime }\right)
=\left\langle \varphi _{2}\right| \varrho ^{\prime }\left| \varphi
_{2}\right\rangle +\left\langle \psi _{2}\right| \varrho ^{\prime }\left|
\psi _{2}\right\rangle   \notag \\
&=&\gamma \left| \varphi _{1}\right\rangle \left\langle \varphi _{1}\right|
+\left( 1-\gamma \right) \left| \psi _{1}\right\rangle \left\langle \psi
_{1}\right|   \notag \\
&=&\frac{1}{2}\left( I+k\chi \sigma _{x}\sin 2\theta -k\sigma _{z}\cos
2\theta \right) ,  \label{e24}
\end{eqnarray}%
where $k=2\gamma -1$. Thus,%
\begin{equation}
d\tilde{\Pi}\left( \chi \right) =\frac{1}{2\pi }\left( I+k\chi \sigma
_{x}\sin 2\theta -k\sigma _{z}\cos 2\theta \right) d\theta ,  \label{e25}
\end{equation}%
and%
\begin{equation}
\tilde{W}\left( \chi \right) =\frac{1}{4\pi }\left( I+k\chi \sigma _{x}\sin
2\theta -k\sigma _{z}\cos 2\theta \right) .  \label{e26}
\end{equation}%
From Eq.(\ref{e13}), we have%
\begin{equation}
\tilde{\Upsilon}=\frac{I}{4\pi }\left[ 1+\frac{1}{2}k^{2}\left( \chi
^{2}+1\right) \right] .  \label{e27}
\end{equation}%
So%
\begin{eqnarray}
&&\tilde{\Upsilon}-\tilde{W}\left( \chi \right)   \notag \\
&=&\frac{1}{8\pi }\left[ k^{2}\left( \chi ^{2}+1\right) -2k\chi \sigma
_{x}\sin 2\theta +2k\sigma _{z}\cos 2\theta \right] ,  \label{e28}
\end{eqnarray}%
which has eigenvalues%
\begin{eqnarray}
\tilde{\lambda} &=&k^{2}\left( \chi ^{2}+1\right) \pm 2k\sqrt{\cos
^{2}2\theta +\chi ^{2}\sin ^{2}2\theta }.  \notag \\
&&  \label{e29}
\end{eqnarray}%
It is clear that if and only if $\theta =\pm \pi /4,$ $\pm 3\pi /4$ we have%
\begin{equation}
\left\{ 
\begin{array}{c}
\left[ \tilde{\Upsilon}-\tilde{W}\left( \chi \right) \right] d\tilde{\Pi}%
\left( \chi \right) =0, \\ 
\tilde{\Upsilon}-\tilde{W}\left( \chi \right) \geqslant 0.%
\end{array}%
\right.   \label{e30}
\end{equation}%
Which shows that it is optimal that the equal probability Schr\"{o}dinger
cat states $\left| \varphi \right\rangle =\left( \left| \alpha \right\rangle
+\left| -\alpha \right\rangle \right) /\sqrt{2}$ and $\left| \psi
\right\rangle =\left( \left| \alpha \right\rangle -\left| -\alpha
\right\rangle \right) /\sqrt{2}$ are taken as the bases of the input states
for the estimation of the mixed lossy channel.

Now we investigate the double lossy channel. Similarly, we take the input
state $\left| \Phi \right\rangle $ as Eq.(\ref{e20}) and the bases as Eq.(%
\ref{e21}). Thus, the initial state $\varrho $ passing through the double
lossy channel becomes%
\begin{eqnarray}
\varrho ^{\prime \prime } &=&\varepsilon \otimes \varepsilon \left( \varrho
\right)   \notag \\
&=&\gamma \varepsilon \left( \left| \varphi _{1}\right\rangle \left\langle
\varphi _{1}\right| \right) \otimes \varepsilon \left( \left| \varphi
_{2}\right\rangle \left\langle \varphi _{2}\right| \right)   \notag \\
&&+\sqrt{\gamma \left( 1-\gamma \right) }\varepsilon \left( \left| \varphi
_{1}\right\rangle \left\langle \psi _{1}\right| \right) \otimes \varepsilon
\left( \left| \varphi _{2}\right\rangle \left\langle \psi _{2}\right|
\right)   \notag \\
&&+\sqrt{\gamma \left( 1-\gamma \right) }\varepsilon \left( \left| \psi
_{1}\right\rangle \left\langle \varphi _{1}\right| \right) \otimes
\varepsilon \left( \left| \psi _{2}\right\rangle \left\langle \varphi
_{2}\right| \right)   \notag \\
&&+\left( 1-\gamma \right) \varepsilon \left( \left| \psi _{1}\right\rangle
\left\langle \psi _{1}\right| \right) \otimes \varepsilon \left( \left| \psi
_{2}\right\rangle \left\langle \psi _{2}\right| \right) .  \notag \\
&&  \label{e31}
\end{eqnarray}%
Because of%
\begin{eqnarray}
\left\langle \varphi _{2}\right| \varepsilon \left( \left| \varphi
_{2}\right\rangle \left\langle \varphi _{2}\right| \right) \left| \varphi
_{2}\right\rangle  &=&1+\frac{1}{2}\left( \chi -1\right) \sin ^{2}2\theta , 
\notag \\
\left\langle \varphi _{2}\right| \varepsilon \left( \left| \psi
_{2}\right\rangle \left\langle \psi _{2}\right| \right) \left| \varphi
_{2}\right\rangle  &=&-\frac{1}{2}\left( \chi -1\right) \sin ^{2}2\theta , 
\notag \\
\left\langle \varphi _{2}\right| \varepsilon \left( \left| \varphi
_{2}\right\rangle \left\langle \psi _{2}\right| \right) \left| \varphi
_{2}\right\rangle  &=&\frac{1}{2}\left( \chi -1\right) \sin 2\theta \cos
2\theta ,  \notag \\
\left\langle \varphi _{2}\right| \varepsilon \left( \left| \psi
_{2}\right\rangle \left\langle \varphi _{2}\right| \right) \left| \varphi
_{2}\right\rangle  &=&\frac{1}{2}\left( \chi -1\right) \sin 2\theta \cos
2\theta ,  \notag \\
&&  \label{e32}
\end{eqnarray}%
and 
\begin{eqnarray}
\left\langle \psi _{2}\right| \varepsilon \left( \left| \varphi
_{2}\right\rangle \left\langle \varphi _{2}\right| \right) \left| \psi
_{2}\right\rangle  &=&-\frac{1}{2}\left( \chi -1\right) \sin ^{2}2\theta , 
\notag \\
\left\langle \psi _{2}\right| \varepsilon \left( \left| \psi
_{2}\right\rangle \left\langle \psi _{2}\right| \right) \left| \psi
_{2}\right\rangle  &=&1+\frac{1}{2}\left( \chi -1\right) \sin ^{2}2\theta , 
\notag \\
\left\langle \psi _{2}\right| \varepsilon \left( \left| \varphi
_{2}\right\rangle \left\langle \psi _{2}\right| \right) \left| \psi
_{2}\right\rangle  &=&-\frac{1}{2}\left( \chi -1\right) \sin 2\theta \cos
2\theta ,  \notag \\
\left\langle \psi _{2}\right| \varepsilon \left( \left| \psi
_{2}\right\rangle \left\langle \varphi _{2}\right| \right) \left| \psi
_{2}\right\rangle  &=&-\frac{1}{2}\left( \chi -1\right) \sin 2\theta \cos
2\theta ,  \notag \\
&&  \label{e33}
\end{eqnarray}%
we have%
\begin{eqnarray}
\varrho _{1}^{\prime \prime } &=&tr_{2}\left( \varrho ^{\prime \prime
}\right) =\left\langle \varphi _{2}\right| \varrho ^{\prime \prime }\left|
\varphi _{2}\right\rangle +\left\langle \psi _{2}\right| \varrho ^{\prime
\prime }\left| \psi _{2}\right\rangle   \notag \\
&=&\gamma \left| \varphi _{1}\right\rangle \left\langle \varphi _{1}\right|
+\left( 1-\gamma \right) \left| \psi _{1}\right\rangle \left\langle \psi
_{1}\right|   \notag \\
&=&\frac{1}{2}\left( I+k\chi \sigma _{x}\sin 2\theta -k\sigma _{z}\cos
2\theta \right) ,  \notag \\
&&  \label{e34}
\end{eqnarray}%
which is similar to $\varrho _{1}^{\prime }$. So we can obtain the similar
results.

\subsection{Composite lossy channels (joint measurements)}

For some purposes jointly measure the output state exported from the
composite channels is needed. The schematic diagram of the joint measurement
is shown in Fig.2.\FRAME{fhFU}{238.875pt}{69.375pt}{0pt}{\Qcb{Schematic
diagram for estimating the composite lossy channel where joint measurement
of output states is supposed}}{}{Figure}{\special{language "Scientific
Word";type "GRAPHIC";maintain-aspect-ratio TRUE;display "USEDEF";valid_file
"T";width 238.875pt;height 69.375pt;depth 0pt;original-width
235.625pt;original-height 67pt;cropleft "0";croptop "1";cropright
"1";cropbottom "0";tempfilename 'HOWX6H06.wmf';tempfile-properties "XPR";}}
In this case, what are the optimal bases of the input states for estimating
the composite lossy channel?\ In this subsection we shall answer this
question.

We use the projector $\Pi =\left| \Psi \right\rangle \left\langle \Psi
\right| $ to measure the output state exported from the composite channels.
Here, we suppose%
\begin{equation}
\left| \Psi \right\rangle =\frac{1}{\sqrt{2}}\left( \left| \varphi
\right\rangle \left| \varphi \right\rangle +\left| \psi \right\rangle \left|
\psi \right\rangle \right) .  \label{eq35}
\end{equation}%
Measuring the the output state $\varrho ^{\prime }$ with projector $\Pi $ we
can obtain its probability as 
\begin{eqnarray}
p^{\prime } &=&\left\langle \Psi \right| \varrho ^{\prime }\left| \Psi
\right\rangle   \notag \\
&=&\frac{1}{2}\left[ \left( \left\langle \varphi \right| \left\langle
\varphi \right| +\left\langle \psi \right| \left\langle \psi \right| \right)
\varrho ^{\prime }\left( \left| \varphi \right\rangle \left| \varphi
\right\rangle +\left| \psi \right\rangle \left| \psi \right\rangle \right) %
\right]   \notag \\
&=&\frac{1}{2}\left\{ 1+\sqrt{\gamma \left( 1-\gamma \right) }\left[ \sin
^{2}\theta +2\gamma \left( \sin ^{4}\theta +\cos ^{4}\theta \right) \right]
\right\} .  \notag \\
&&  \label{eq36}
\end{eqnarray}%
Similarly, measuring the output state $\varrho ^{\prime \prime }$ with
projector $\Pi $ we can obtain its probability as

\begin{eqnarray}
p^{\prime \prime } &=&\left\langle \Psi \right| \varrho ^{\prime \prime
}\left| \Psi \right\rangle  \notag \\
&=&\frac{1}{2}\left[ \left( \left\langle \varphi \right| \left\langle
\varphi \right| +\left\langle \psi \right| \left\langle \psi \right| \right)
\varrho ^{\prime \prime }\left( \left| \varphi \right\rangle \left| \varphi
\right\rangle +\left| \psi \right\rangle \left| \psi \right\rangle \right) %
\right]  \notag \\
&=&\frac{1}{2}\left\{ 1+\sqrt{\gamma \left( 1-\gamma \right) }\left[ \sin
^{2}\theta +2\gamma \left( \sin ^{4}\theta +\cos ^{4}\theta \right) \right]
\right\} .  \notag \\
&&  \label{eq37}
\end{eqnarray}%
It shows that when $\theta =\pm \pi /4,$ the probabilities $p^{\prime }$ and 
$p^{\prime \prime }$ take their maximum.

In this section we have obtained that the equal probability Schr\"{o}dinger
cat states are optimal for estimating the single lossy channel and they are
also the optimal bases of input states for estimating the composite lossy
channels.

\section{SLD Fisher information of the output states of lossy channels}

Based on the above results, in this section we shall investigate the second
problem put forward in section II, namely, can the entangled input states
improve the estimation of the composite lossy channels? To accomplish this
task we may use another method, to calculate the SLD Fisher information of
the output states of the lossy channels. At first, we briefly review this
theory. Given a one-parameter family $\rho \left( \varsigma \right) $ of
density operator, an estimator for parameter $\varsigma $ is represented by
a Hermitian operator $T$. It is shown that if the system is in the state $%
\rho _{\varsigma },$ then the expectation $E_{\varsigma }\left[ T\right]
:=Tr\rho _{\varsigma }T$ of the estimator $T$ should be identical to $%
\varsigma $ and the estimator $T$ for the parameter $\varsigma $ satisfies
the quantum Cram\'{e}r-Rao inequality $V_{\varsigma }\left[ T\right]
\geqslant \left( J_{\varsigma }\right) ^{-1},$ where $V_{\varsigma }\left[ T%
\right] :=Tr\rho _{\varsigma }\left( T-\varsigma \right) ^{2}$ is the
variance of estimator $T,$ and $J_{\varsigma }:=J\left( \rho _{\varsigma
}\right) :=Tr\rho _{\varsigma }\left( L_{\varsigma }\right) ^{2}$ is the
quantum SLD Fisher information with $L_{\varsigma }$ the symmetric
logarithmic derivative. Here, the Hermitian operator $L_{\varsigma }$
satisfies the equation%
\begin{equation}
\frac{d\rho _{\varsigma }}{d\varsigma }=\frac{1}{2}\left( L_{\varsigma }\rho
_{\varsigma }+\rho _{\varsigma }L_{\varsigma }\right) .  \label{eq38}
\end{equation}%
It is important to notice that the lower bound $\left( J_{\varsigma }\right)
^{-1}$ in the quantum Cram\'{e}r-Rao inequality is achievable (at lest
locally). In other words, the inverse of the SLD Fisher information gives
the ultimate limit of estimation. So in our problem the bigger of the SLD
Fisher information is, the more accurately the estimation may be \cite%
{Fujiwara042304}$.$ In \cite{Fujiwara042304} the author has proved that the
SLD Fisher information is convex so we only need to investigate the pure
state inputs. In the following we will calculate the SLD Fisher information
of output states for above three lossy channels.

\subsection{Optimal inputs cases}

In the previous section we obtain that the equal probability Schr\"{o}dinger
cat states are the optimal input states for estimating the single lossy
channel, and they are also the optimal bases of the input states for
estimating the composite channels. In the following we calculate the SLD
Fisher information of the output state for above three lossy channels by use
of optimal input states or optimal bases of the input states.

At first, we calculate the SLD Fisher information of single lossy channel
when the input state is $\rho =\left| \varphi \right\rangle \left\langle
\varphi \right| .$ In this problem, the parameter $\varsigma $ in above
formulas is $\chi $ in the output states of lossy channels$.$ From Eq.(\ref%
{e9}) we have%
\begin{equation}
\frac{d\rho ^{\prime }}{d\chi }=\left( 
\begin{array}{cc}
0 & -\left| \alpha \right| ^{2}\tau \chi  \\ 
-\left| \alpha \right| ^{2}\tau \chi  & 0%
\end{array}%
\right) ,  \label{eq39}
\end{equation}%
and 
\begin{equation}
L_{\chi }=\frac{2\left| \alpha \right| ^{2}\chi }{1-\chi ^{2}}\left( 
\begin{array}{cc}
\chi  & -1 \\ 
-1 & \chi 
\end{array}%
\right) .  \label{eq40}
\end{equation}%
So we can easily calculate the SLD Fisher information of output state $\rho
^{\prime }$ as 
\begin{equation}
J_{\chi }^{\varepsilon }\left| _{\theta =\frac{\pi }{4}}\right. =\frac{%
4\left| \alpha \right| ^{4}\tau ^{2}e^{-4\left| \alpha \right| ^{2}\eta \tau
}}{1-e^{-4\left| \alpha \right| ^{2}\eta \tau }}.  \label{eq41}
\end{equation}

Secondly, we calculate the SLD Fisher information of the mixed lossy channel
by using the input state $\varrho =\left| \Phi \right\rangle \left\langle
\Phi \right| $ (see Eq.(\ref{e20})). Here 
\begin{eqnarray}
\left| \varphi _{1}\right\rangle  &=&\left| \varphi _{2}\right\rangle =\frac{%
1}{\sqrt{2}}\left( \left| \alpha \right\rangle +\left| -\alpha \right\rangle
\right) ,  \notag \\
\left| \psi _{1}\right\rangle  &=&\left| \psi _{2}\right\rangle =\frac{1}{%
\sqrt{2}}\left( \left| \alpha \right\rangle -\left| -\alpha \right\rangle
\right) .  \label{eq42}
\end{eqnarray}%
Thus, we have 
\begin{equation}
\rho _{\chi }^{\prime }\left| _{\theta =\frac{\pi }{4}}\right. =\varepsilon
\otimes I\left( \varrho \right) =\frac{1}{4}\left( 
\begin{array}{cccc}
u & w & w\chi  & u\chi  \\ 
w & v & v\chi  & w\chi  \\ 
w\chi  & v\chi  & v & w \\ 
u\chi  & w\chi  & w & u%
\end{array}%
\right) ,  \label{eq43}
\end{equation}%
where $u=1+2\sqrt{\gamma \left( 1-\gamma \right) },$ $v=1-2\sqrt{\gamma
\left( 1-\gamma \right) },$ $w=2\gamma -1.$ Here, the expression of
Hermitian operator $L_{\chi }$ is too complex, we do not give it.
Fortunately, by using Eq.(\ref{eq43}) we can obtain a simple expression of
its SLD Fisher information as%
\begin{equation}
J_{\chi }^{\varepsilon \otimes I}\left| _{\theta =\frac{\pi }{4}}\right. =%
\frac{4\left| \alpha \right| ^{4}\tau ^{2}e^{-4\left| \alpha \right|
^{2}\eta \tau }}{1-e^{-4\left| \alpha \right| ^{2}\eta \tau }}.  \label{eq44}
\end{equation}%
It is shown that the SLD Fisher information of mixed channel do not vary
with the changing of the entanglement degree of input state, namely, it does
not vary with $\gamma ^{\prime }s$ changing. The SLD Fisher information of
mixed lossy channel is accurately equal to one of one-shot single lossy
channel. It means that when we take the equal probability Schr\"{o}dinger
cat states as the bases the entangled input states the estimation of the
mixed lossy channel can not be improved.

Thirdly, we investigate the double lossy channel. By using the input state $%
\varrho =\left| \Phi \right\rangle \left\langle \Phi \right| $ we have the
output state as%
\begin{equation}
\begin{tabular}{l}
$\varrho _{\chi }^{\prime \prime }\left| _{\theta =\pi /4}\right.
=\varepsilon \otimes \varepsilon \left( \varrho \right) $ \\ 
$\qquad =\left[ 
\begin{array}{cccc}
\frac{1+2A}{4} & \frac{\chi B}{4} & \frac{\chi B}{4} & \frac{\chi ^{2}\left(
2A+1\right) }{4} \\ 
\frac{\chi B}{4} & \frac{1-2A}{4} & \frac{\chi ^{2}\left( 1-2A\right) }{4} & 
\frac{\chi B}{4} \\ 
\frac{\chi B}{4} & \frac{\chi ^{2}\left( 1-2A\right) }{4} & \frac{1-2A}{4} & 
\frac{\chi B}{4} \\ 
\frac{\chi ^{2}\left( 2A-1\right) }{4} & \frac{\chi B}{4} & \frac{\chi B}{4}
& \frac{1+2A}{4}%
\end{array}%
\right] $%
\end{tabular}%
,  \label{eq45}
\end{equation}%
where, $A=\sqrt{\gamma \left( 1-\gamma \right) },$ $B=2\gamma -1.$ By using
the $\varrho _{\chi }^{\prime \prime }\left| _{\theta =\pi /4}\right. $ we
can calculate its SLD Fisher information $J_{\chi }^{\varepsilon \otimes
\varepsilon }\left| _{\theta =\pi /4}\right. .$ But the expression of $%
J_{\chi }^{\varepsilon \otimes \varepsilon }\left| _{\theta =\pi /4}\right. $
is too complex and too long. Here, we do not give it, too. We plot the $%
J_{\chi }^{\varepsilon \otimes \varepsilon }\left| _{\theta =\pi /4}\right. $
with $\gamma $ and time $\tau $ as Fig.3 where $\eta =0.25$ and $\left|
\alpha \right| =3.$ \FRAME{ftbpFU}{221.875pt}{148.125pt}{0pt}{\Qcb{SLD
Fisher information $J_{\protect\chi }^{\protect\varepsilon \otimes \protect%
\varepsilon }\left| _{\protect\theta =\protect\pi /4}\right. $ of double
lossy channel with $\protect\gamma $ and time $\protect\tau ,$ where $\left| 
\protect\alpha \right| =3,$ $\protect\eta =0.25.$}}{}{Figure}{\special%
{language "Scientific Word";type "GRAPHIC";display "USEDEF";valid_file
"T";width 221.875pt;height 148.125pt;depth 0pt;original-width
216.8125pt;original-height 160.375pt;cropleft "0";croptop "1";cropright
"1";cropbottom "0";tempfilename 'HOWX6H09.wmf';tempfile-properties "XPR";}}%
From Fig.3, we see that when we take the equal probability Schr\"{o}dinger
cat states as the bases of the input states, the entangled inputs can not
improve the estimation of the double lossy channel, too.

Before we end this subsection we analytically investigate two kinds of
specific cases. Namely, we calculate the SLD Fisher information of output
state of double lossy channel when the input states are product state and
maximally entangled state where optimal bases is still held. When the input
state is a product state, namely, $\gamma =0$ or $\gamma =1$, the density
operator of the output state is%
\begin{equation}
\varrho _{\chi }^{\prime \prime }\left| _{\substack{ \gamma =0,1 \\ \theta
=\pi /4}}\right. =\varepsilon \otimes \varepsilon \left( \varrho \right) =%
\frac{1}{4}\left( 
\begin{array}{cccc}
1 & \pm \chi  & \pm \chi  & \chi ^{2} \\ 
\pm \chi  & 1 & \chi ^{2} & \pm \chi  \\ 
\pm \chi  & \chi ^{2} & 1 & \pm \chi  \\ 
\chi ^{2} & \pm \chi  & \pm \chi  & 1%
\end{array}%
\right) ,  \label{eq46}
\end{equation}%
where ``+'' denotes a product state of $\gamma =1$, and ``-'' denotes a
product state of $\gamma =0$. Thus, we have 
\begin{equation}
L_{\chi }^{\varepsilon \otimes \varepsilon }\left| _{\substack{ \gamma =0,1
\\ \theta =\pi /4}}\right. =\frac{2\left| \alpha \right| ^{2}\tau \chi }{%
1-\chi ^{2}}\left( 
\begin{array}{cccc}
2\chi  & 1 & 1 & 0 \\ 
1 & 2\chi  & 0 & 1 \\ 
1 & 0 & 2\chi  & 1 \\ 
0 & 1 & 1 & 2\chi 
\end{array}%
\right) .  \label{eq47}
\end{equation}%
So when the input state is the product state, the SLD Fisher information of
double lossy channel is%
\begin{equation}
J_{\chi }^{\varepsilon \otimes \varepsilon }\left| _{\substack{ \gamma =0,1
\\ \theta =\frac{\pi }{4}}}\right. =\frac{8\left| \alpha \right| ^{4}\tau
^{2}e^{-4\left| \alpha \right| ^{2}\eta \tau }}{1-e^{-4\left| \alpha \right|
^{2}\eta \tau }}.  \label{eq48}
\end{equation}%
This result is reasonable, because $\gamma =0$ or $\gamma =1$ and $%
\varepsilon \otimes \varepsilon \left( \varrho \right) $ is correspond to
the input of product state of two-shot single lossy channel and the result
Eq.(\ref{eq48}) is just the two times of the SLD Fisher information of
single lossy channel.

When we set $\gamma =\frac{1}{2}$ the input state $\left| \Phi \right\rangle 
$ is the maximally entangled state and we can obtain 
\begin{equation}
\varrho _{\chi }^{\prime \prime }\left| _{\substack{ \gamma =1/2 \\ \theta
=\pi /4}}\right. =\varepsilon \otimes \varepsilon \left( \rho \right)
=\left( 
\begin{array}{cccc}
\frac{1}{2} & 0 & 0 & \frac{1}{2}\chi ^{2} \\ 
0 & 0 & 0 & 0 \\ 
0 & 0 & 0 & 0 \\ 
\frac{1}{2}\chi ^{2} & 0 & 0 & \frac{1}{2}%
\end{array}%
\right) ,  \label{eq49}
\end{equation}%
and 
\begin{equation}
L^{\varepsilon \otimes \varepsilon }\left| _{\substack{ \gamma =1/2 \\ %
\theta =\pi /4}}\right. =\frac{4\left| \alpha \right| ^{2}\tau \chi ^{2}}{%
1-\chi ^{4}}\left( 
\begin{array}{cccc}
\chi ^{2} & 0 & 0 & -1 \\ 
0 & L_{22}^{\prime } & L_{23}^{\prime } & 0 \\ 
0 & L_{32}^{\prime } & L_{33}^{\prime } & 0 \\ 
-1 & 0 & 0 & \chi ^{2}%
\end{array}%
\right) ,  \label{eq50}
\end{equation}%
where $L_{22}^{\prime },$ $L_{23}^{\prime },$ $L_{32}^{\prime },$ and $%
L_{33}^{\prime }$ are uncertain matrix elements. Fortunately, by using the
uncertain $L^{\varepsilon \otimes \varepsilon }\left| _{\gamma =1/2,\theta
=\pi /4}\right. ,$ we can calculate the SLD Fisher information as 
\begin{equation}
J_{\chi }^{\varepsilon \otimes \varepsilon }\left| _{\substack{ \gamma =1/2
\\ \theta =\pi /4}}\right. =\frac{16\left| \alpha \right| ^{4}\tau
^{2}e^{-8\left| \alpha \right| ^{2}\eta \tau }}{1-e^{-8\left| \alpha \right|
^{2}\eta \tau }}.  \label{eq51}
\end{equation}%
This is the SLD Fisher information of channel $\varepsilon \otimes
\varepsilon $ when the input state is the maximally entangled state. A
numerical work shows that $J_{\chi }^{\varepsilon \otimes \varepsilon
}\left| _{\gamma =1/2,\theta =\pi /4}\right. $ is less than $J_{\chi
}^{\varepsilon \otimes \varepsilon }\left| _{\gamma =0or1,\theta =\pi
/4}\right. $ for $\left| \alpha \right| >1.$

In this subsection we obtained that when we take the equal probability
Schr\"{o}dinger cat states as the bases of the input states the entangled
input states can not improve the estimation of composite lossy channels.

\subsection{Non-optimal inputs cases}

In this subsection, we discuss another case. When the input states are some
non-optimal states, namely, when $\theta \neq \pi /4$ in the input states,
whether the entangled inputs are better than the non entangled inputs for
estimating these composite lossy channels? In the following, we only discuss
the cases of $\theta =0,$ $\pi /2.$ Thus, In this cases the input state Eq.(%
\ref{e20}) becomes 
\begin{equation}
\left| \Phi \right\rangle =\sqrt{\gamma }\left| \alpha \right\rangle \left|
\alpha \right\rangle +\sqrt{1-\gamma }\left| -\alpha \right\rangle \left|
-\alpha \right\rangle .  \label{eq52}
\end{equation}%
For the mixed channel, the output state is%
\begin{equation}
\varrho _{\chi }^{\prime }\left| _{\substack{ \theta =0, \\ \theta =\pi /2}}%
\right. =\left[ 
\begin{array}{cccc}
\gamma  & 0 & 0 & \sqrt{\gamma \left( 1-\gamma \right) } \\ 
0 & 0 & 0 & 0 \\ 
0 & 0 & 0 & 0 \\ 
\sqrt{\gamma \left( 1-\gamma \right) } & 0 & 0 & 1-\gamma 
\end{array}%
\right] .  \label{eq53}
\end{equation}%
Thus we can calculate the SLD Fisher information of $\varrho _{\chi
}^{\prime }\left| _{_{\theta =0,\theta =\pi /2}}\right. $ as%
\begin{eqnarray}
&&J_{\chi }^{\varepsilon \otimes I}\left| _{_{\substack{ \theta =0, \\ %
\theta =\pi /2}}}\right.   \notag \\
&=&\frac{16\gamma \left( \gamma -1\right) \chi ^{4}\tau ^{2}\left| \alpha
\right| ^{4}}{4\gamma \left( 1-\gamma \right) \gamma ^{4}+\left( 8\gamma
^{2}-8\gamma +1\right) \gamma ^{2}-(2\gamma -1)^{2}}.  \notag \\
&&  \label{eq54}
\end{eqnarray}%
Here, we do not give out the SLD operator for its very complex. Similarly,
we can obtain the output state of the double lossy channel as 
\begin{equation}
\varrho _{\chi }^{\prime \prime }\left| _{_{\substack{ \theta =0, \\ \theta
=\pi /2}}}\right. =\left[ 
\begin{array}{cccc}
\gamma  & 0 & 0 & \sqrt{\gamma \left( 1-\gamma \right) }\chi ^{2} \\ 
0 & 0 & 0 & 0 \\ 
0 & 0 & 0 & 0 \\ 
\sqrt{\gamma \left( 1-\gamma \right) }\chi ^{2} & 0 & 0 & 1-\gamma 
\end{array}%
\right] .  \label{eq55}
\end{equation}%
Its SLD Fisher information is%
\begin{eqnarray}
&&J_{\chi }^{\varepsilon \otimes \varepsilon }\left| _{_{\substack{ \theta
=0, \\ \theta =\pi /2}}}\right.   \notag \\
&=&\frac{64\gamma \left( \gamma -1\right) \chi ^{8}\tau ^{2}\left| \alpha
\right| ^{4}}{4\gamma \left( 1-\gamma \right) \chi ^{4}+\left( 8\gamma
^{2}-8\gamma +1\right) \chi ^{2}-(2\gamma -1)^{2}}.  \notag \\
&&  \label{eq56}
\end{eqnarray}%
Here, we also do not give the the SLD operator for its very complex. From
Eqs.(\ref{eq55}), (\ref{eq56}) we can easily obtain that $J_{\chi
}^{\varepsilon \otimes I}\left| _{_{\theta =0,\theta =\pi /2}}\right. $ and $%
J_{\chi }^{\varepsilon \otimes \varepsilon }\left| _{_{\theta =0,\theta =\pi
/2}}\right. $ take their maximum at $\gamma =1/2$ (see Figs.4 and 5), which
shows that when the bases of input states are not optimal bases( equal
probability Schr\"{o}dinger cat states) the entangled input states may
improve the estimation of the composite lossy channels.\FRAME{fhFU}{251.875pt%
}{126pt}{0pt}{\Qcb{SLD Fisher information $J_{\protect\chi }^{\protect%
\varepsilon \otimes I}\left| _{\protect\theta =0,\protect\theta =\protect\pi %
/2}\right. $ of mixed lossy channel with $\protect\gamma $ and time $\protect%
\tau ,$ where $\protect\eta =0.25,$ $\protect\alpha =3.$}}{}{Figure}{\special%
{language "Scientific Word";type "GRAPHIC";maintain-aspect-ratio
TRUE;display "USEDEF";valid_file "T";width 251.875pt;height 126pt;depth
0pt;original-width 339.5625pt;original-height 171.25pt;cropleft "0";croptop
"0.9658";cropright "0.9806";cropbottom "0";tempfilename
'HOWX6H07.wmf';tempfile-properties "XPR";}}\FRAME{fhFU}{251.875pt}{126pt}{0pt%
}{\Qcb{SLD Fisher information $J_{\protect\chi }^{\protect\varepsilon %
\otimes \protect\varepsilon }\left| _{\protect\theta =0,\protect\theta =%
\protect\pi /2}\right. $ of double lossy channel with $\protect\gamma $ and
time $\protect\tau ,$ where $\protect\eta =0.25,$ $\protect\alpha =3.$}}{}{%
Figure}{\special{language "Scientific Word";type
"GRAPHIC";maintain-aspect-ratio TRUE;display "USEDEF";valid_file "T";width
251.875pt;height 126pt;depth 0pt;original-width 341.0625pt;original-height
171.25pt;cropleft "0";croptop "0.9730";cropright "0.9835";cropbottom
"0";tempfilename 'HOWX6H08.wmf';tempfile-properties "XPR";}}

\section{Conclusions}

In this paper, we have discussed two corresponding problems to estimating
quantum lossy channels. Firstly, we have investigated what the optimal input
states is for estimating the single lossy channel and what the best bases
are of the input states for estimating the composite lossy channels. We
obtain that the equal probability Schr\"{o}dinger cat states are the optimal
input states for estimating the single lossy channel and they are also the
optimal bases of input states for estimating the composite lossy channels.
Secondly, we have investigated that whether the entangled input states can
improve the estimation of the quantum lossy channels. By calculating the SLD
Fisher information of the output states we obtained that when we take the
equal probability Schr\"{o}dinger states as the bases of the input states
the entangled input states can not improve the estimation of the composite
channel, however when we take the coherent states $\left| \pm \alpha
\right\rangle $ instead of the equal probability Schr\"{o}dinger cat states
as the bases, the maximally entangled input states can improve the
estimation of the composite lossy channels.

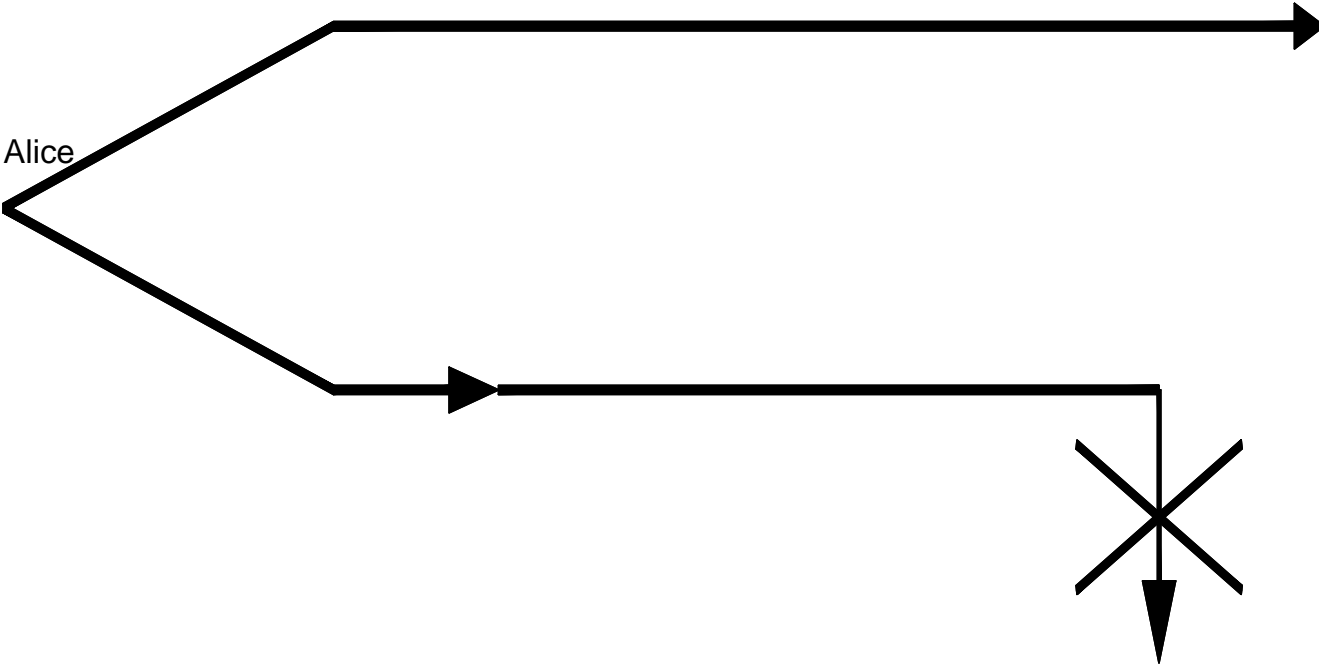
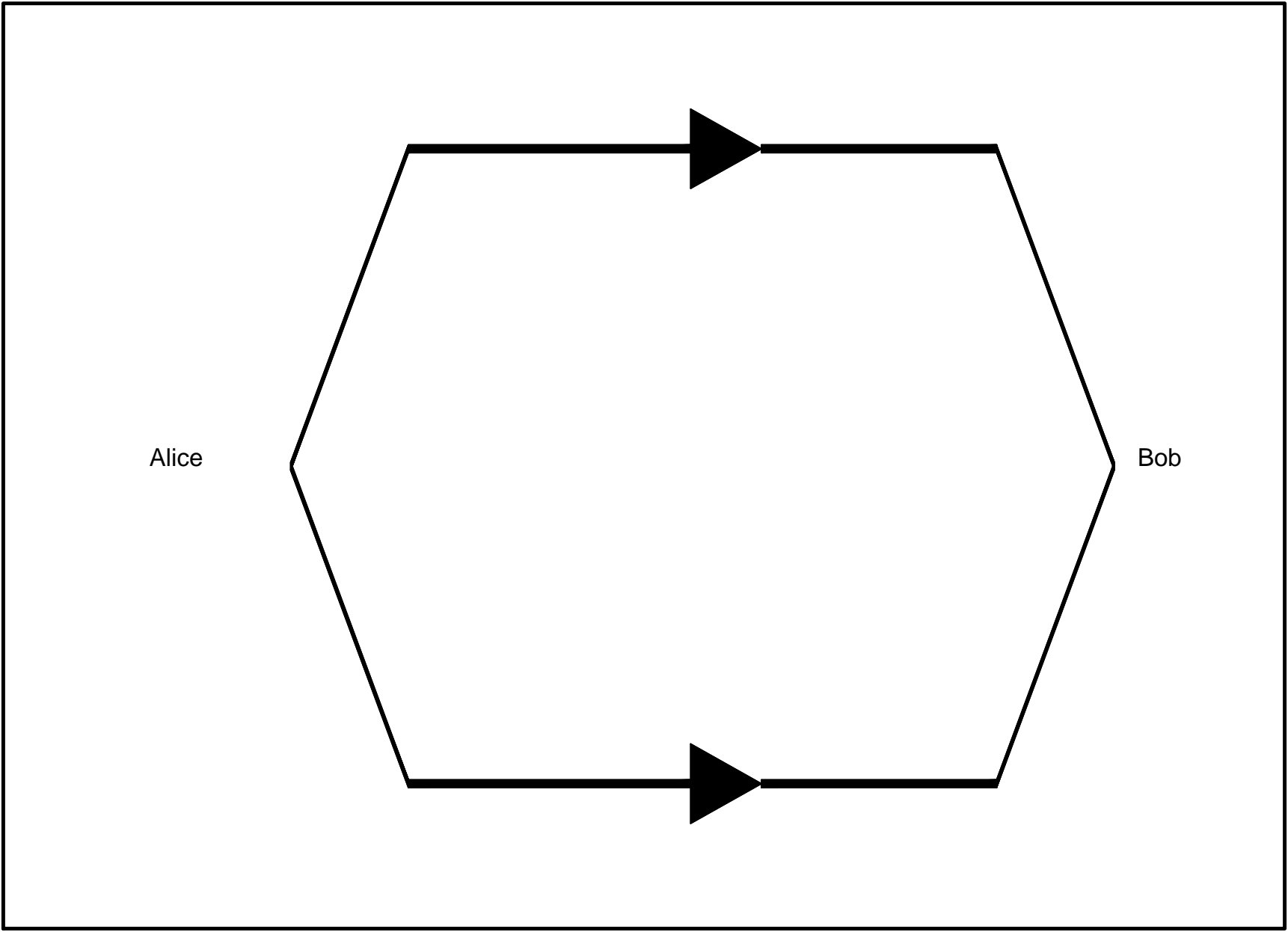
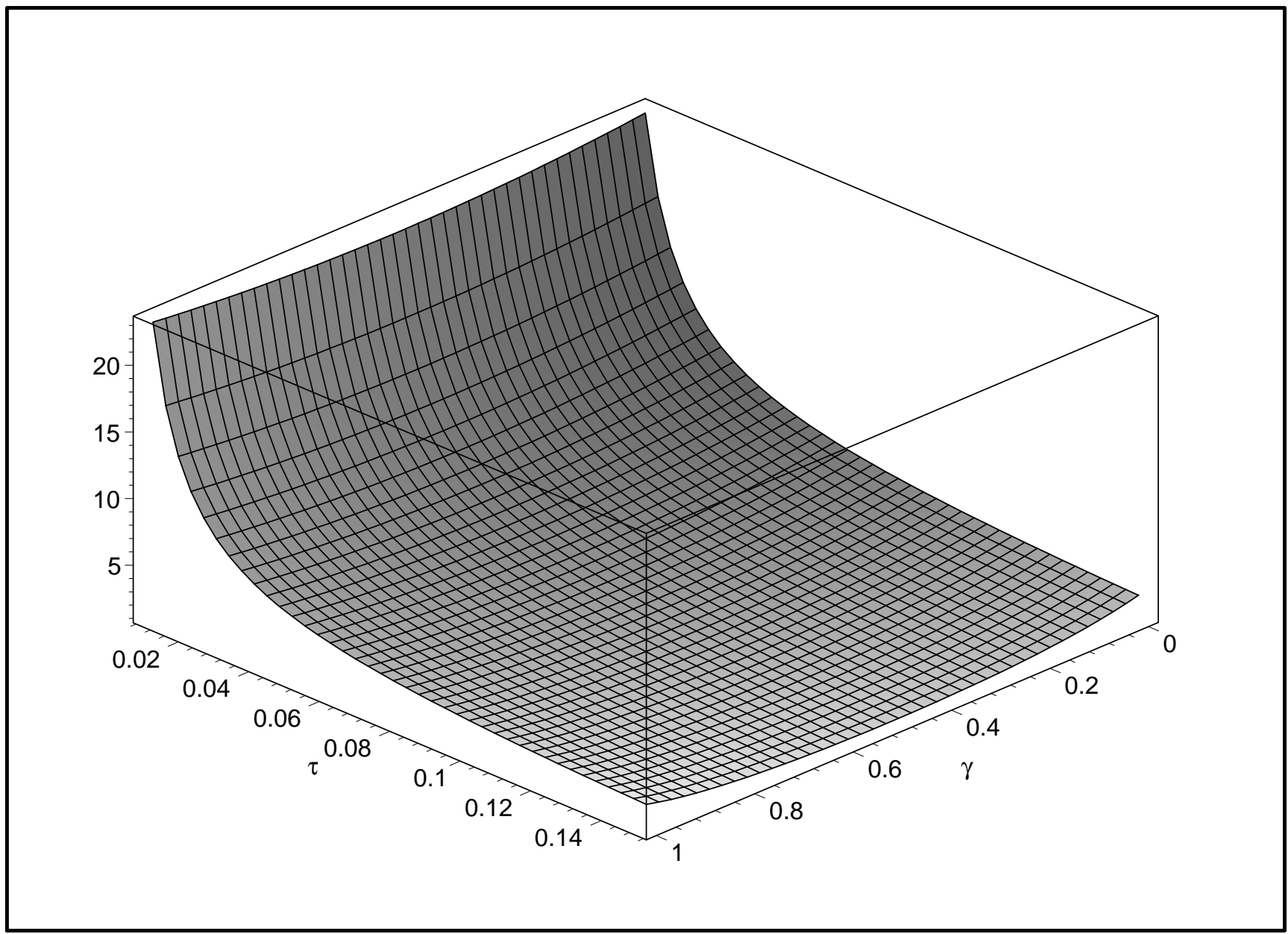
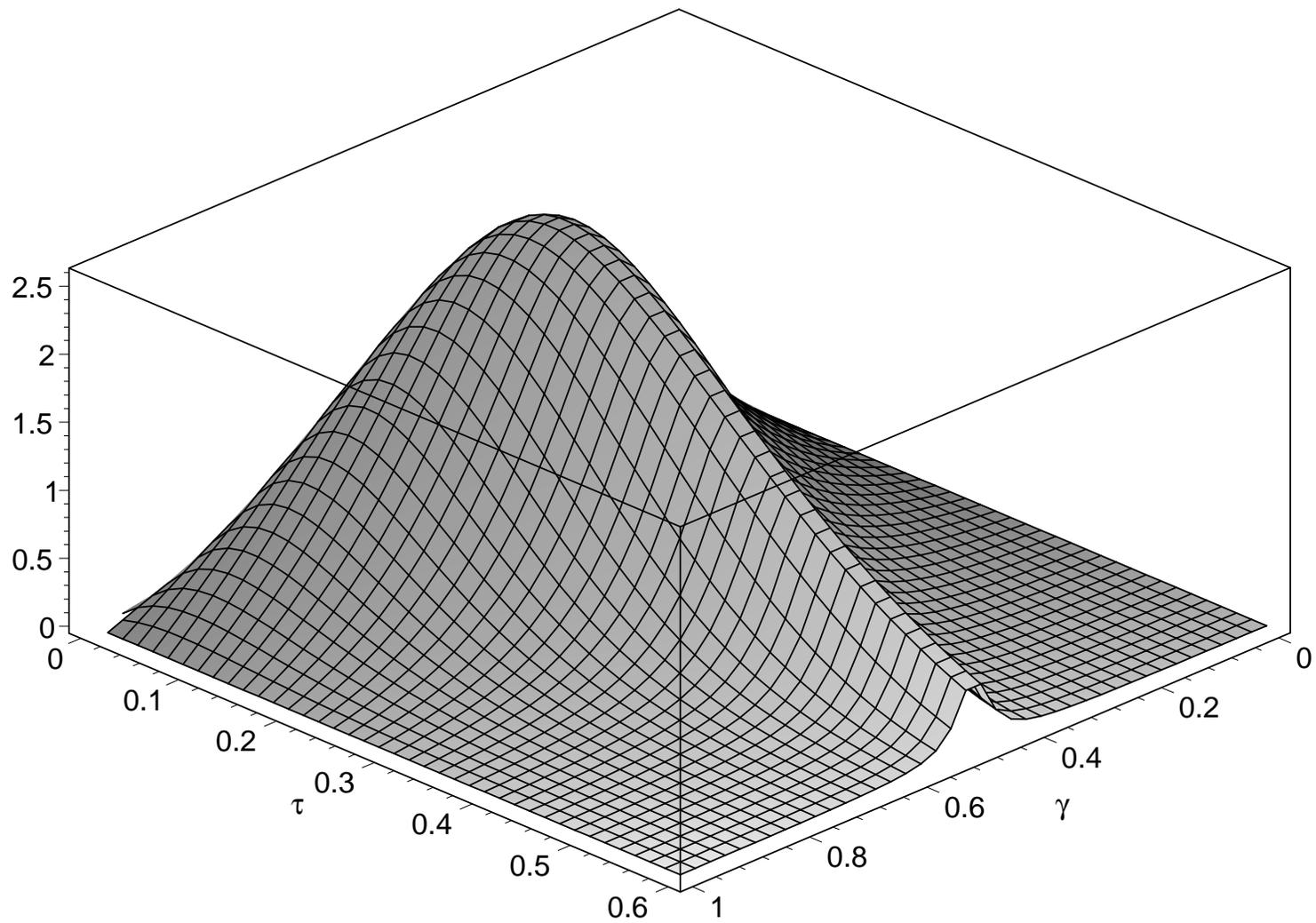
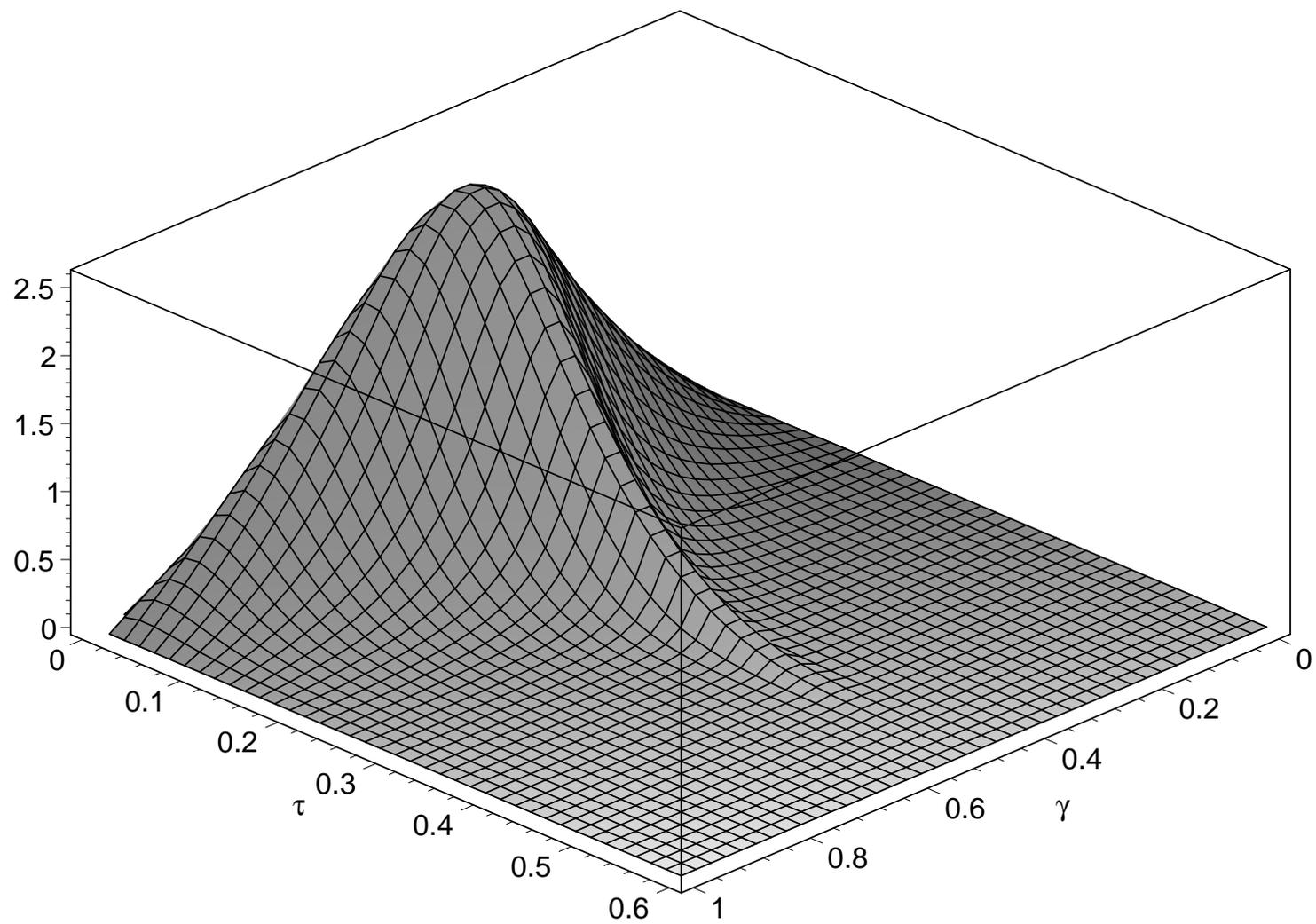
\end{document}